\documentclass[12pt,psfig]{article}

\begin{document}

\title {Glass transition in Ultrathin 
Polymer Films : A Thermal Expansion Study}

\author {M. Bhattacharya, M. K. Sanyal\\
Surface Physics Division, Saha Institute of Nuclear Physics,\\
1/AF, Bidhannagar, Kolkata 700 064, India.\\
Th. Geue, U. Pietsch\\
Institut f\"{v}r Physik, Universit\"{a}t Potsdam,\\
D-14469 Potsdam, Germany.}
\date{\today}

\maketitle

\begin{abstract}
Glass transition process gets affected in ultrathin films having 
thickness comparable to the size of the molecules. We observe 
systematic broadening of glass transition temperature ($T_g$) as 
the thickness of the polymer film reduces below the radius of 
gyration but the change in the average $T_g$ was 
found to be very small. Existence of reversible 
negative and positive thermal expansion below and above $T_g$ 
increased the sensitivity of our thickness measurements performed 
using energy dispersive x-ray reflectivity. 
A simple model of $T_g$ variation as a function of depth expected from 
sliding motion could explain the results. We observe clear glass 
transition even for 4 nm polystyrene film that was predicted to be 
absent from ellipsometry measurements of thicker films.

\end{abstract}

\section{Introduction}

The phase transition process of non-crystalline materials from a 
glassy to molten state has remained an outstanding problem in 
condensed matter physics especially when these materials are 
confined in nanometer length scale. This is a subject of intense 
research for macromolecules like polymers as the confinement effect 
becomes prominent even in films having thickness of tens of nanometers. 
The effect of nano-confinement on the glass transition temperature 
($T_g$) of ultrathin polymer films is being studied extensively 
\cite{reiter01,forrest02} as the results may 
influence our basic understanding about glass transition mechanism 
in general \cite{keddie94,wallace95} and dynamics of polymer chains 
near glass-melt 
transition, in particular \cite{reiter93,degennes00}. 
These studies are also important 
for technological applications of nanometer-thick polymer films 
\cite{ellison04}. 
In most of these investigations the thicknesses of polymer films are 
measured using ellipsometry technique \cite{kawana01} as a 
function of temperature across glass-melt transition. The 
$T_g$ in these measurements was defined \cite{forrest02,kawana01} using 
temperatures at which thermal expansion deviates from 
linearity above ($T_+$) and below ($T_-$) the glass transition with 
$T_g$ = 0.5 ($T_+ + T_-$). In ellipsometry measurements, $T_+$ 
were found to be constant and $T_-$ exhibited rapid reduction 
as the thicknesses of the polymer films approached 10 nm | 
a thickness below which $T_g$ could not be 
determined from thickness measurements due to contrast problem 
\cite{forrest02,keddie94,kawana01}. As $T_+$ was constant, both broadening 
of glass transition 
$\Delta T$ (= $T_+ - T_-$) and $T_g$ itself were found to follow the 
variation of $T_-$ with thickness in polystyrene 
thin films. It was assumed that for thin polymer films a top layer of 
thickness around 5 nm is in molten state even at room temperature.
The effect of the molten top free surface on reduction of $T_g$ has been 
confirmed by capping the polymer films by a metal layer \cite{sharp03}.
The measurements were performed on primarily polystyrene films having 
thickness between 10 to 100 nanometer and the values 
of $T_g$ thus obtained from various different measurements could be 
parameterized with an empirical relation 
\begin{equation}
T_g^f = T_g^b [1- (A/h)^\delta]
\end{equation} 
where $T_g^b$ and $T_g^f$ are glass transition temperatures for bulk 
polymer and for a film of thickness {\it h} respectively 
\cite{forrest02,keddie94,kawana01}. This equation obviously indicates 
finite $T_g$ above a thickness A and its value was found to be 3.2 nm. 
It is indeed very interesting to note that except one data set 
\cite{wallace95}, large number of measurements 
\cite{forrest02,kawana01,sharp03} performed in different experimental 
conditions with variety of surface treatments of silicon substrates, 
which were used for depositing polystyrene films of various molecular 
weights, followed this empirical relation with A = 3.2 nm and 
$\delta$ = 1.8.
The implication of this intriguing result is that a film of 3.2 nm will 
not show finite $T_g$ and even for a 8 nm film $T_g^f$ will be below 
room temperature. Our results however exhibit evidence of glass transition 
for films having even 4 nm thickness and the obtained value of 
$T_g^f$ remains close to $T_g^b$ though 
transition becomes broad as the thickness is reduced. 
We also demonstrate that one has to use the concept of distributed glass 
transition temperature \cite{degennes00} to explain the dramatic 
increase in $\Delta T$ as the thickness becomes comparable to the 
radius of gyration ($R_g$) of the polymer.
In a recent calorimetric measurement, ultrathin (1 to 10nm)
polymer films exhibited \cite{efremov03} pronounced glass transition 
temperature and this measurement also showed minor change in $T_g$ with 
thickness but confirmed the broadening of glass transition with 
reduction of film thickness, which is in agreement with our 
findings. Moreover, the results presented here are consistent with 
the earlier thickness measurement studies using conventional x-ray 
reflectivity and ellipsometry techniques except for the films less 
than 20 nm for which sensitivity of earlier thickness measurements 
were not satisfactory \cite{forrest02,keddie94,kawana01}.

\section{Experimental Studies}

\subsection{Sample preparation}
We have studied thermal expansion behaviour of spin coated ultrathin 
Poly-styrene films in the thickness range of 4 to 31 nm. 
Polystyrene (Aldrich, USA, $M_w$ = 212,400, radius of gyration 
$R_g$$\sim$12.6 nm and $T_g^b$ = 373K) thin films were deposited 
by spin coating 
solutions of the polymer in toluene at around 3000 rpm onto 
hydrophilic silicon (100) substrates prepared following 
RCA cleaning procedure. In this chemical treatment, substrates
were first heated with 30\% ammonium hydroxide ($NH_4OH$) 
and 30\% hydrogen peroxide ($H_2O_2$) 
solution (1:1 volume ratio) for 10 minutes followed by a rinsing 
with acetone and alchohol and finally washed with purifies 
water having 18.2 Mohm cm resistivity (Millipore, USA). The film 
thickness was controlled by changing the rotation speed and by 
choosing appropriate concentration of the solution 
from 1-6.5 mg/ml of toluene. 

\subsection{Sample characterization}

It is known that any property of polymer films determined through 
thickness measurements is sensitive to dewetting phenomena and 
one has to be careful especially for 
ultrathin polymer films \cite{reiter93,tolan98,basu99,zhao93}. 
We present here results of polymer films deposited on hydrophilic 
silicon (100) substrate, which do not show any dewetting even after 
repeated thermal cycling. The polymer films were heated in a sample 
cell under vacuum (3$\times$ 10$^{-2}$ mbar) at a temperature (403K) 
above $T_g^b$ of PS and we have monitored the 
top surface of the film with optical microscope. We have performed 
these measurements for both thick and ultrathin PS films down to 
4.3 nm range reported here. In Fig. 1(a), the obtained ADR profiles
and corresponding fits are presented as a function of temperature for
a 4.3 nm thick polymer film. From our analysis, it is seen that 
the roughness values of this film of 
4.3 nm thickness measured at 303 K and 403 K come out to be 
2.8 \AA~ and 3.3 \AA~ respectively (Fig. 1(a)) and we could get back 
same parameter after repeated thermal cycling. Dewetting of film 
would have increased the roughness drastically \cite{reiter93} and we 
could not have got back same reflectivity profile after thermal cycling.

We have also performed Energy Dispersive X-ray Spectrometry measurements 
on chemically treated silicon substrate and the polymer films. In the 
inset of Fig. 1(a), part of EDX spectra of a substrate and polymer film 
collected at 5 keV energy in scanning electron microscope are shown to 
indicate that the film is free from other contaminants and our 
impression is that clean environment should be used to avoid dewetting 
of the polystyrene films deposited on hydrophilic silicon substrates.

\subsection{Transverse x-ray diffuse scattering measurements}

We have studied the in-plane correlation present in polymer films by 
measuring transverse x-ray diffuse scattering as a function 
of temperature at X22C beamline of National Synchrotron Light Source, 
Brookhaven National Laboratory. In Fig. 1(b) and (c) we have shown data 
collected at two temperatures as a function of 
$q_x [= (2 \pi /\lambda) (cos \beta - cos \alpha)]$ at three fixed 
$q_z [= (2 \pi /\lambda) (sin \alpha + sin \beta)]$ values, where 
$\alpha$ and $\beta$ are incidence and exit angle of x-rays of wavelength 
$\lambda$. Capillary waves are of particular importance in case 
of soft matter interfaces and polymer films are known 
\cite{tolan98,zhao93} to
exhibit charactaristic of capillary wave correlation
\cite{tolan98,basu99,zhao93,daillant00,sanyal91}. 
From theory of capillary waves, 
the height-height correlation function can be expressed as 
\begin{equation}
C(R)=(B/2) K_0 (\kappa R),
\end{equation}

where $K_0 (\kappa R)$ is modified Bessel function, $\kappa$ is 
low wave vector cut off and $B = k_B T/(\pi \gamma)$ with $\gamma$
as the surface tension at temperature T and $k_B$ = 1.3806 $\times$ 
10$^{-23}$ J/K is the Boltzmann constant. Within experimentally 
accessible wave vector range, 
$K_0 (\kappa R)$ can be well approximated as 
$[K_0 (\kappa R) \approx -ln (\kappa R/2)-\gamma_E]$,
where $\gamma_E$ is the Euler 
constant. Inserting this type of logarithmic correlation to the 
scattering function calculation and considering an approximation 
of the transmission functions $|T(\alpha)|^2 |T(\beta)|^2$ with 
that of substrate the observed intensity can be 
obtained as \cite{sanyal91},

\begin{eqnarray}
I = I_0 \frac{R(q_z)q_z}{2k_0 sin\alpha} \frac{1}{\sqrt \pi} 
~exp[-q_z^2\sigma_e^2] ~~ \Gamma\left(\frac{1-\eta}{2}\right)\nonumber \\ 
\times _1F_1 \left(\frac{1-\eta}{2};\frac{1}{2};\frac{q_x^2 L^2}{4\pi^2}
\right) |T(\alpha)|^2 |T(\beta)|^2 
\end{eqnarray}

Here $_1 F_1 (x;y;z)$ is the Kummar function and we could fit all the 
data using Eq. (3) self consistently \cite{sanyal91} (Fig. 1(b) and (c)) 
with this line shape. From our analysis, the values of 
B come out to be 1.1 \AA~$^{2}$ and 2.1 \AA~$^{2}$ 
for 333 K and 393 K respectively. Dewetting of films would have 
generated different line shapes as observed earlier \cite{zhao93}.
We find that the films retain this line shape through thermal 
cycling confirming that dewetting is not taking place.

\subsection{Thickness measurements through reflectivity}

Thermal expansion studies of the polymer thin films were carried 
out in the temperature range of 308 K to 433 K using both energy 
(EDR) and angle (ADR) dispersive 
x-ray reflectivity techniques \cite{mukherjee02}.
In specular condition, the intensity of reflected 
x-ray beam is measured as a function of wave vector transfer 
$q_z ( = 4\pi E \sin \theta /(hc) $ with hc = 12.3986 keV \AA~) where 
the incident and exit angle $\theta$ are kept equal. The ADR 
experiments were performed by changing the angle $\theta$ 
in a laboratory set up \cite{basu99} and the EDR measurements were 
carried out at the EDR beamline \cite{bodenthin02} of BESSY-II 
synchrotron (Berlin, Germany) by keeping the angle 
$\theta$ constant (1$^o$ here). All the reflectivity measurements 
both in angle and energy dispersive mode were carried out 
keeping the samples in vacuum (10$^{-2}$ mbar).

We have already demonstrated \cite{mukherjee02} that the polymer 
films deposited on 
hydrophilic Si(100) substrate exhibit reversible negative and positive 
thermal expansion below and above $T_g^b$ respectively. Our observation has 
been confirmed in an independent measurement and predicted large 
relaxation time has also been measured \cite {miyazaki04}. 
Negative thermal expansion 
was also observed in polycarbonate ultrathin film while cooling 
the films below apparent $T_g$ \cite{christopher04}. 
The existence of positive and negative thermal expansion \cite{mukherjee02,miyazaki04, christopher04, orts93} coefficients 
above and below $T_g$ 
improved the sensitivity of measurement of glass transition temperature. 
Moreover, we used here the EDR technique, which allowed us to collect 
large number of reflectivity profiles around $T_g^b$ in quick succession.
EDR technique is ideal method to investigate small thermal expansion 
in thin polymer films. 
In Fig. 2 we have shown the measured ADR and EDR profiles collected 
from a polymer film at room temperature along with the fit 
giving us a thickness of 4.5 nm. The procedure of EDR data extraction 
and other experimental details of EDR beamline of BESSY-II synchrotron 
has been described earlier \cite{mukherjee02,bhattacharya03}. 
We have also shown some of the 
representative EDR profiles at three different temperatures along 
with the fitted curves in the inset of Fig. 2. 
The change in thickness is apparent in the shift of minimum of 
reflectivity curves itself. In the analysis \cite{bhattacharya03} 
of normalized EDR and ADR data, we have used 
a model of a polymer film of constant electron density with two 
roughness profiles at film-air and film-substrate interfaces. 
The obtained thickness as a function of temperature for this film is
shown in Fig. 3(a). Initially, 
the 4.5 nm PS film showed negative thermal expansion with the coefficient 
of -3.7$\times$ 10$^{-3}$ K$^{-1}$ over $\sim$30 K temperature range 
starting from 308 K. Near $\sim$339 K, negative thermal expansion 
decreases to give rise to a flat region up to $\sim$408 K in the 
thickness vs. temperature curve. Above $\sim$408 K, the film 
thickness increases and the resultant expansion coefficient comes to 
be 2.5$\times$ 10$^{-3}$ K$^{-1}$, 
much greater than that of bulk PS \cite{mukherjee02,bhattacharya03}. 
The temperatures $T_+$ and $T_-$ were determined by fitting 
two straight lines at high and low temperature region respectively 
as shown in the Fig. 3(a). The calculated $T_g^f$ obtained from 0.5 
($T_+ + T_-$) (indicated as $T_m$) and from the intersection of the 
two fitted lines as shown in Fig. 3(a) were found to be very close 
for all the films. In Fig. 3(b) and (c) we have shown results of 
the films having room temperature thickness of 5.7 nm and 23.1 nm
respectively. We find that values of $T_g^f$ remains very close to 
that for the bulk polymer $T_g^b$ for all films but broadening of 
glass transition 
$\Delta T (=T_+ - T_-)$ reduces as the film thickness increases.

\section{Results and Discussions}

It has been argued \cite{degennes00} that 
for ultrathin films one can expect two types of diffusion
processes, the first one is standard motions controlled by the 
free volume and the second one is sliding ({\it reptation}) 
motions for chains having end points at the free surface. One can 
expect increase in the sliding 
motions for molecules having centre of mass near surface giving rise 
to decrease in $T_g$ towards free surface of a film. In a recent 
fluorescence measurement of multilayer polymer films, systematic 
reduction of $T_g$ was observed for the molecules close to free surface 
\cite{ellison03}. We have used this concept for explaining observed 
thickness variation with temperature and large 
broadening ($\Delta T$) of glass transition.
In our simple model, we have assumed that a film is composed 
of small pseudo-layers of equal thickness having different glass 
transition 
temperatures ($T_g$). Each of these pseudo-layers continues to show 
negative thermal expansion up to the corresponding $T_g$, 
independent of other pseudo-layers and then expands giving rise 
to positive thermal expansion. The negative and positive thermal 
expansion coefficients in each of these pseudo-layers were kept 
equal to that of the total film. The $T_g$ -s, obtained 
from best fit of the measured curve corresponding to the $T_g$ of six 
pseudo-layers used here for the 4.5 nm and 5.7 nm PS films, are 
indicated in the inset of Fig. 3(a) and 3(b). The generated curve by 
adding all these six pairs of straight line match quite well with 
experimental data. The fit can 
be improved by using more number of pseudo-layers but the basic feature 
is quite clear even in this simple model. In case of 23.1 nm film 
(Fig. 3(c)) we used five pseudo-layers instead of six as $\Delta T $ has 
reduced significantly. The molecules near surface have enhanced 
probability of sliding motion and thereby lower effective $T_g$. 
Effective $T_g$ of pseudo-layers progressively approach bulk value 
$T_g^b$ as the position of the pseudo-layer lowers and this value becomes 
even more than $T_g^b$ for the pseudo-layer situated near film substrate 
interface due to strong interaction between polymer and substrate. 
In our model, $T_-$ and $T_+$ are effective $T_g$ of the top and 
bottom pseudo-layers in each film. 
The calculated values of 
$T_g^f$, $T_+$ and $T_-$ for all the films are shown in Fig. 4(a). 
In Fig. 4(b), we have presented obtained absolute value of thermal 
expansion ($|\alpha|$) both below and above 
$T_g$ and the extracted $\Delta T$ for the 
five PS films of different thicknesses. We note that the values of 
$|\alpha|$ in these films decreases exponentially as the film 
thickness is increased and from $\sim$ 2$R_g$-thick PS films, 
the thermal expansion above $T_g$ become close to the volume 
expansion of bulk PS as observed earlier 
\cite{mukherjee02}. 
The broadening of glass transition $\Delta T$ also exhibit an 
exponential dependence and we could parameterized all the curves 
with a function $ a_1 [exp(-h/a_2)]+ a_3 $.
The values of fitting parameters $a_1$, $a_2$, $a_3$ for 
negative and positive thermal expansion coefficients come out to be 
4.1$\times$ 10$^{-2}$ K$^{-1}$, 1.8 nm, 
2.2$\times$ 10$^{-4}$ K$^{-1}$ and 
5.6$\times$ 10$^{-3}$ K$^{-1}$, 
4.4 nm, 4.5$\times$ 10$^{-4}$ K$^{-1}$ respectively. 
For the $\Delta T$ data, we obtain these parameters as 
85.46 K, 10.8 nm and 15.1 K respectively.

In Fig. 4(a) 
we have also shown variation of $T_g^f$ with thickness and 
found that there is a reduction of 7 K as the 
thickness of the film is reduced to 4 nm with an error bar around
3 K. Our results agree with the calorimetric results 
\cite{efremov03} which indicated no appreciable dependence of 
glass-transition temperature on film thickness. In fact one averaged 
value of $T_g$ for the ultrathin films having large $\Delta T$ is 
an ill-defined parameter. Referring back to Eq. (1) and its 
remarkable success in explaining large number of ellipsometry results, 
we note that this equation represents essentially variation of 
$T_-$ as $T_+$ was found to be constant in ellipsometry
measurements.
We extracted the values of $T_-$ as a function of thickness 
from the published values \cite{forrest02,kawana01} of $T_g^f$
knowing the constant values of $T_+$ in those ellipsometry 
measurements. The extracted values of $T_-$ are shown with 
the obtained values of $T_-$ from our measurements 
in the inset of Fig. 4(a). 
Excluding the present data, one can indeed 
get respective values of {\it A} and $\delta$ as 3.76 $\pm$ 0.88 nm 
and 1.76 $\pm$ 0.35, by fitting (dashed line) the Eq. (1) as reported 
\cite{forrest02,keddie94,kawana01}. 
For fitting the data set over the entire thickness range shown in 
inset of Fig. 4(a) with the Eq. (1) one has to ignore the earlier measured 
ellipsometry data below 20 nm that had problem of contrast 
\cite{kawana01}. We obtained the values of 0.47 $\pm$ 0.21 nm 
and 1.13 $\pm$ 0.23 for {\it A} and $\delta$ respectively from 
this fitting (solid line). It may be noted that the obtained 
value of {\it A} here is close to the segment length of 
polystyrene 0.67 nm indicating that polymer film of finite thickness 
will always have finite $T_-$. Moreover increase in $T_+$ for 
ultrathin films will keep $T_g^f$ close to $T_g^b$.

\section{Conclusion}

In conclusion, we have shown by measuring thickness of ultrathin 
polystyrene films as a function of temperature that conventional glass 
transition takes place even for a 4.5 nm film and we do not observe an 
appreciable dependence of average $T_g$ as a function of thickness. 
We find large broadening of glass transition temperature occur as 
the film thickness reduces below radius of gyration. We could explain 
this broadening 
by invoking continuous glass transition temperature as a function of 
depth arising due to variation of the contribution of sliding motion 
in the diffusion process. Probability of having {\it ends} of a polymer 
chain at the free surface, the essential requirement for sliding motion, 
increases as the centre of mass of the chain comes closer to the free 
surface \cite{degennes00} giving rise to lowest but finite $T_g$, 
marked as $T_-$ in a film.

\begin{center}
{\bf ACKNOWLEDGEMENTS}
\end{center}

The authors are grateful to Manabendra Mukherjee and Yves Bodenthin 
for valuable discussions. The authors are also thankful to Joydeep Basu 
and John Hill for their support in NSLS experiment. This work was supported by DST-DAAD India Germany 
collaboration programme.

\newpage
\begin{figure}
{\bf Figure captions:}

\caption{(a) The measured ADR profiles (shifted vertically) 
of a 4.3 nm film with fits are shown as a function 
of temperature, top to bottom, for a thermal cycle.
The shift of minima, marked by dashed lines, indicate 
the reversibility of thermal expansion. In the inset 
part of the EDX spectra collected from a Silicon 
substrate (line) and a ultrathin polymer film (line+circle) 
are shown. 
In (b) and (c), log-log plots of transverse diffuse scattering 
intensity are presented along with the fitted curves 
\cite{sanyal91} (solid lines) as a function of $q_x$ at three 
values of $q_z$ (with different symbols) at 333 K and 393 K
respectively. Also the backgrounds used for fitting are 
indicated by dashed lines.}

\caption{The ADR (circle) and extracted \cite{mukherjee02} 
EDR (square) 
profiles at room temperature for another 4.5 nm PS film are shown 
along with the fitted line. Two dashed lines indicate the usable
EDR data range. Typical EDR profiles for this film at 
temperatures 333K (star), 373 K (up triangle), 429 K (square) and 
321 K (down triangle) are also shown along with the 
fitted profiles in the inset.}

\caption {(a) The obtained thickness variation for the 4.5 nm PS 
film is presented as a function of temperature for heating 
cycle along with the generated curve obtained from our 
model. The temperatures $T_-$, $T_+$, $T_m$, $T_g^f$ are 
indicated and the broadening of glass transition $\Delta T$ is 
marked by straight dashed lines with arrows. Another straight dashed 
line is used to show $T_m$ and $T_g^f$ are very close (refer text 
for details).In the inset of (a) the six straight lines are shown 
indicating the $T_g$-s of six pseudo-layers as obtained from our model 
(the profiles are vertically shifted for clarity). Results of similar 
analysis is shown for films having room temperature thickness of 
5.7 nm (b) and 23.1 nm (c) (refer text for details). }

\caption{(a) $T_+$ (solid up triangle), 
$T_g^f$ (open circle), and $T_-$ (solid down triangle) are presented 
with the error bars as a function of film thickness for five 
PS films of thicknesses 4.5 nm, 5.7 nm, 11.6 nm, 23.1 nm and 31.6 nm. 
In the inset of (a) the solid line represents the fit with the 
Eq. (1) to the $T_-$ data (solid down triangle) of our films 
and for the thicker films published earlier \cite{forrest02}. 
Inclusion of earlier published thinner 
film $T_-$ values (open down triangle) clearly indicate rapid 
reduction (dashed line) of $T_-$ erroneously.
(b) The obtained negative thermal expansion (solid square) 
and positive thermal expansion (open square) coefficients 
together with $\Delta T$ (star) for the five PS films are presented 
with exponential fits (refer text for details).}

\end{figure}
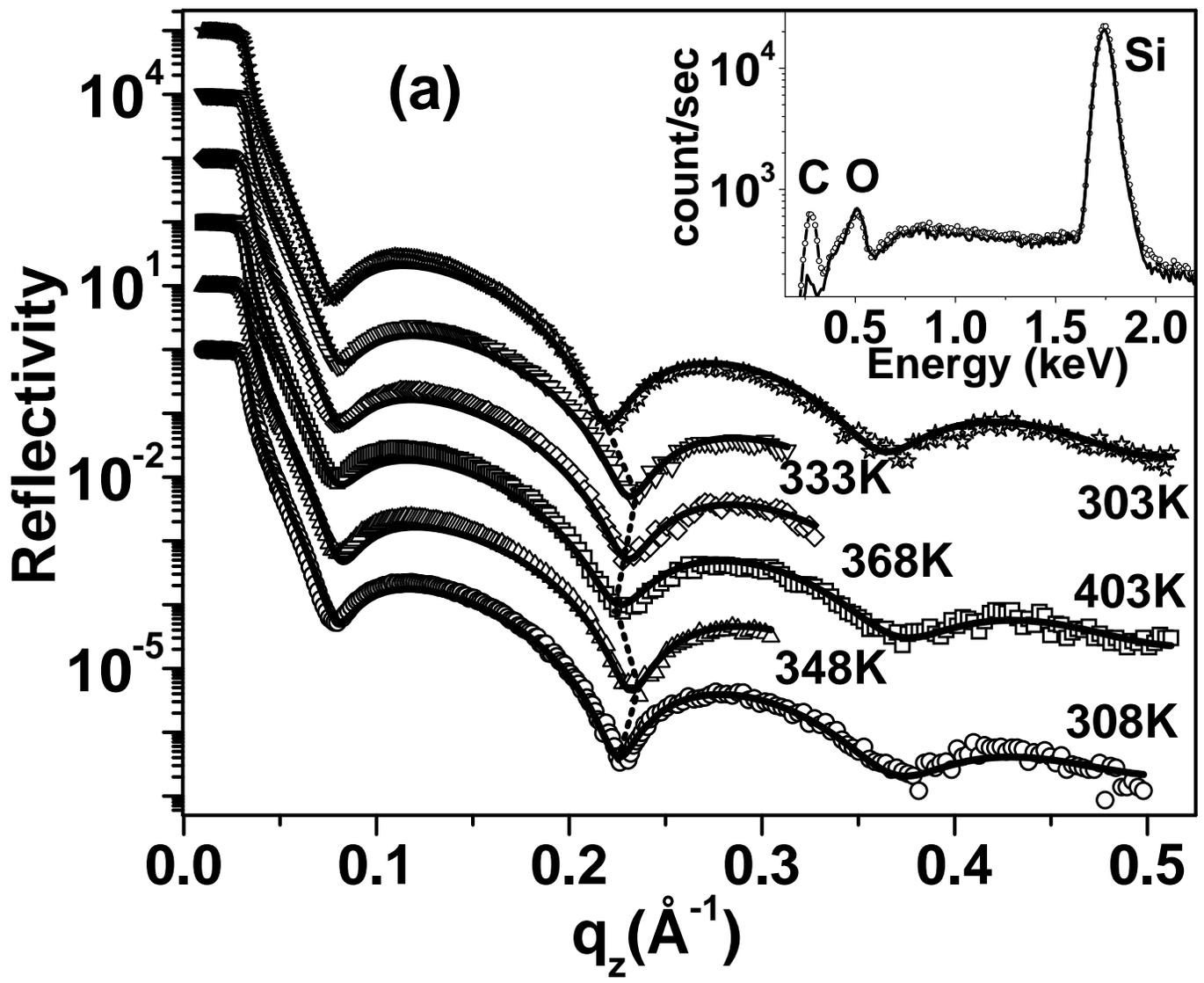
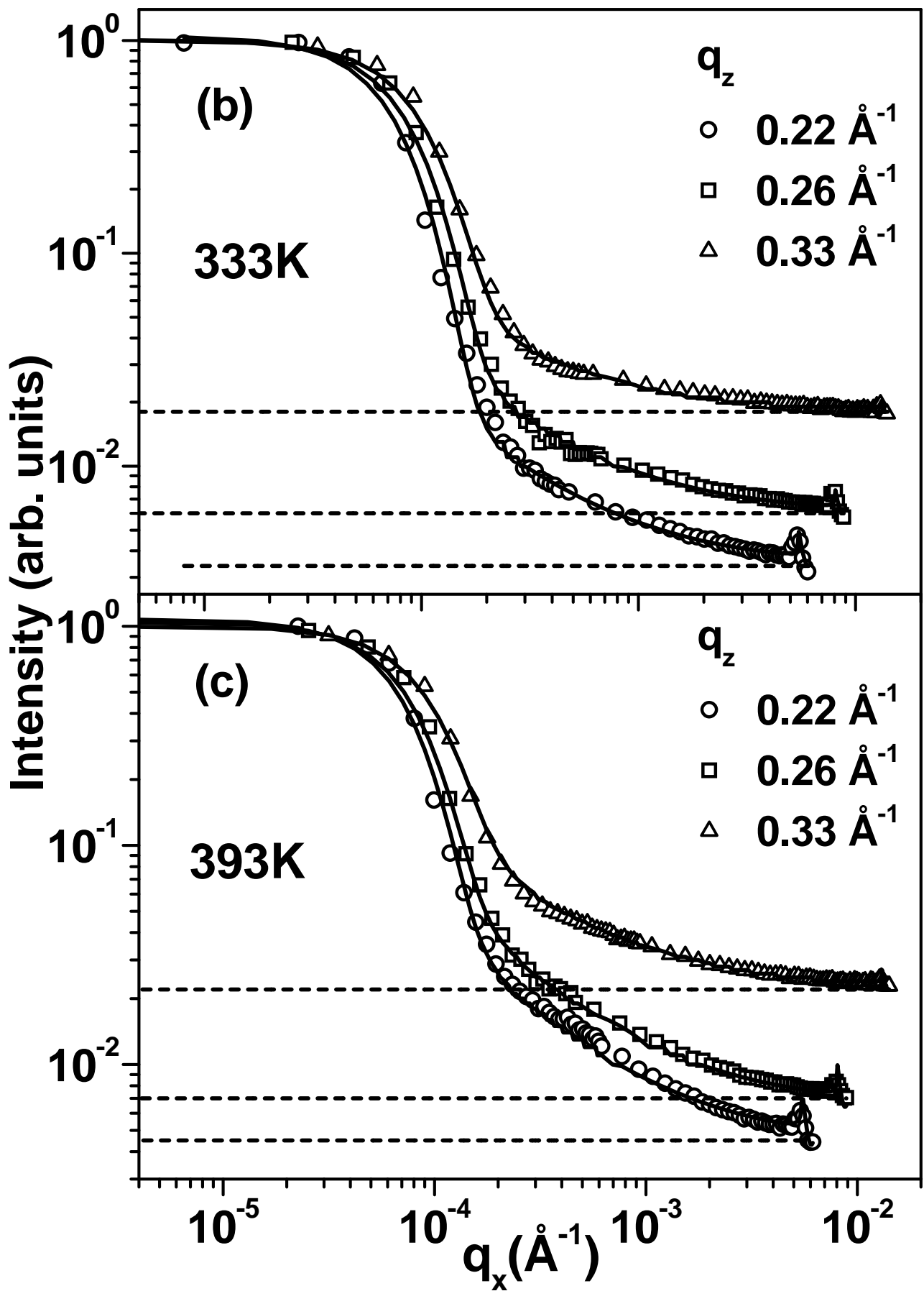
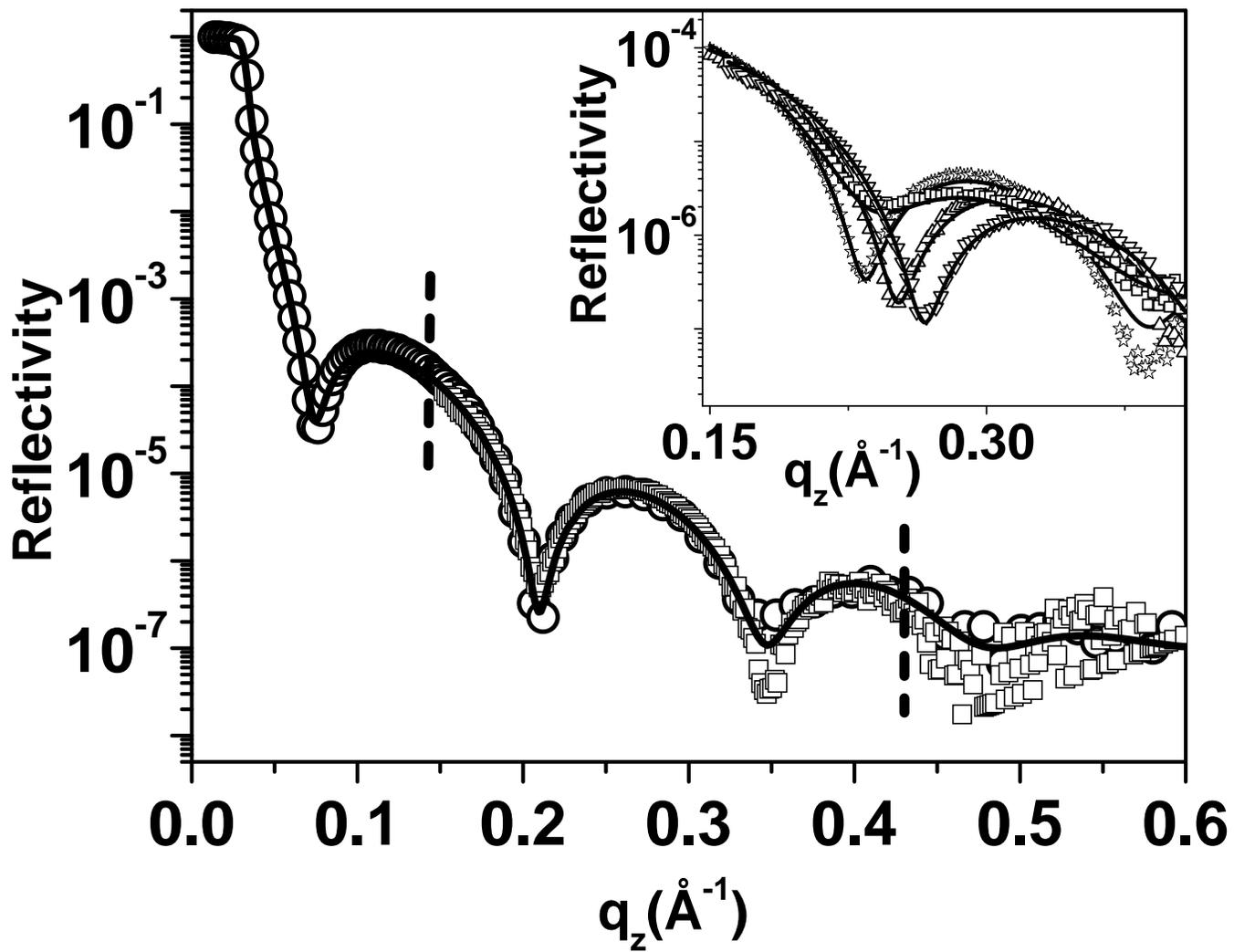
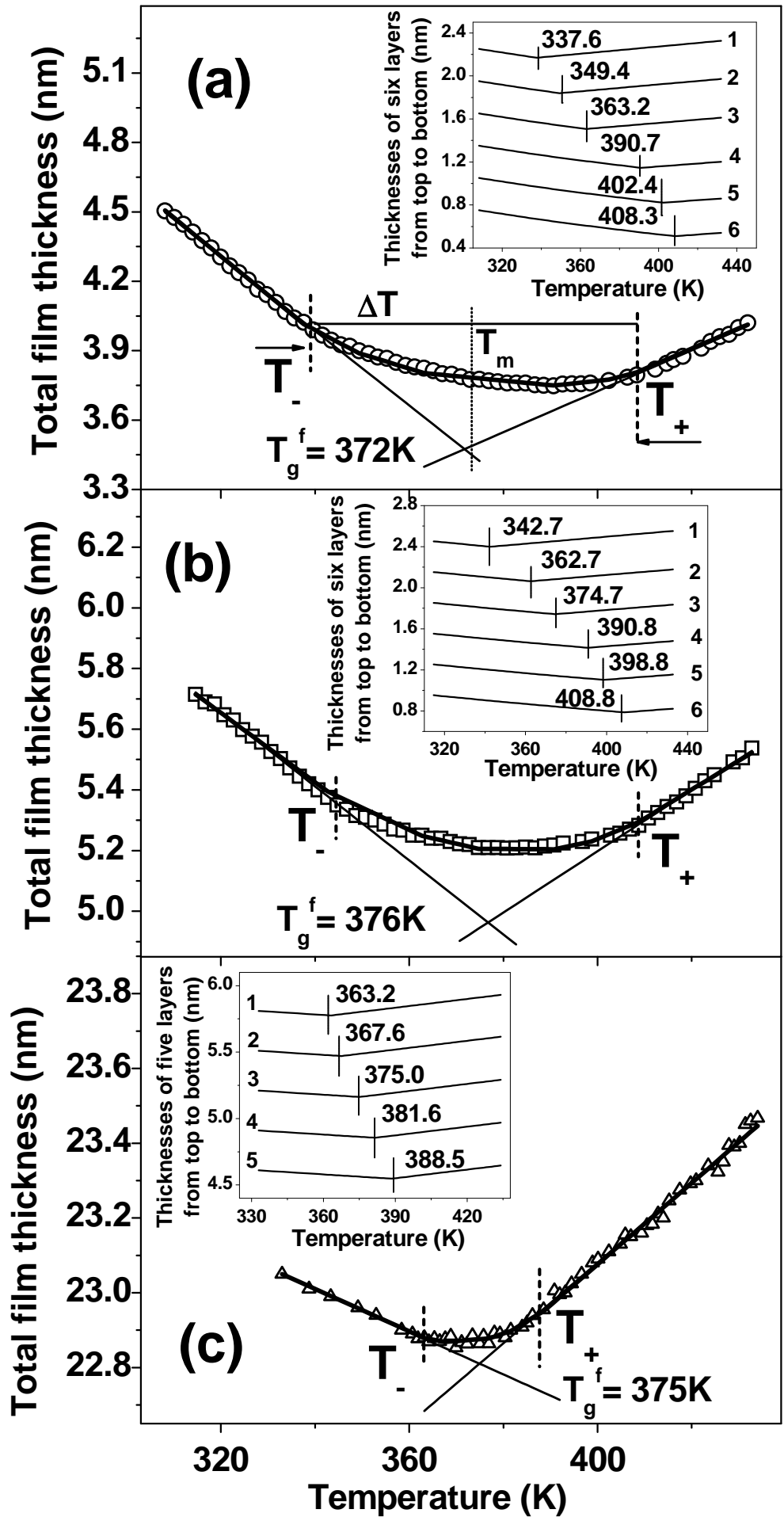
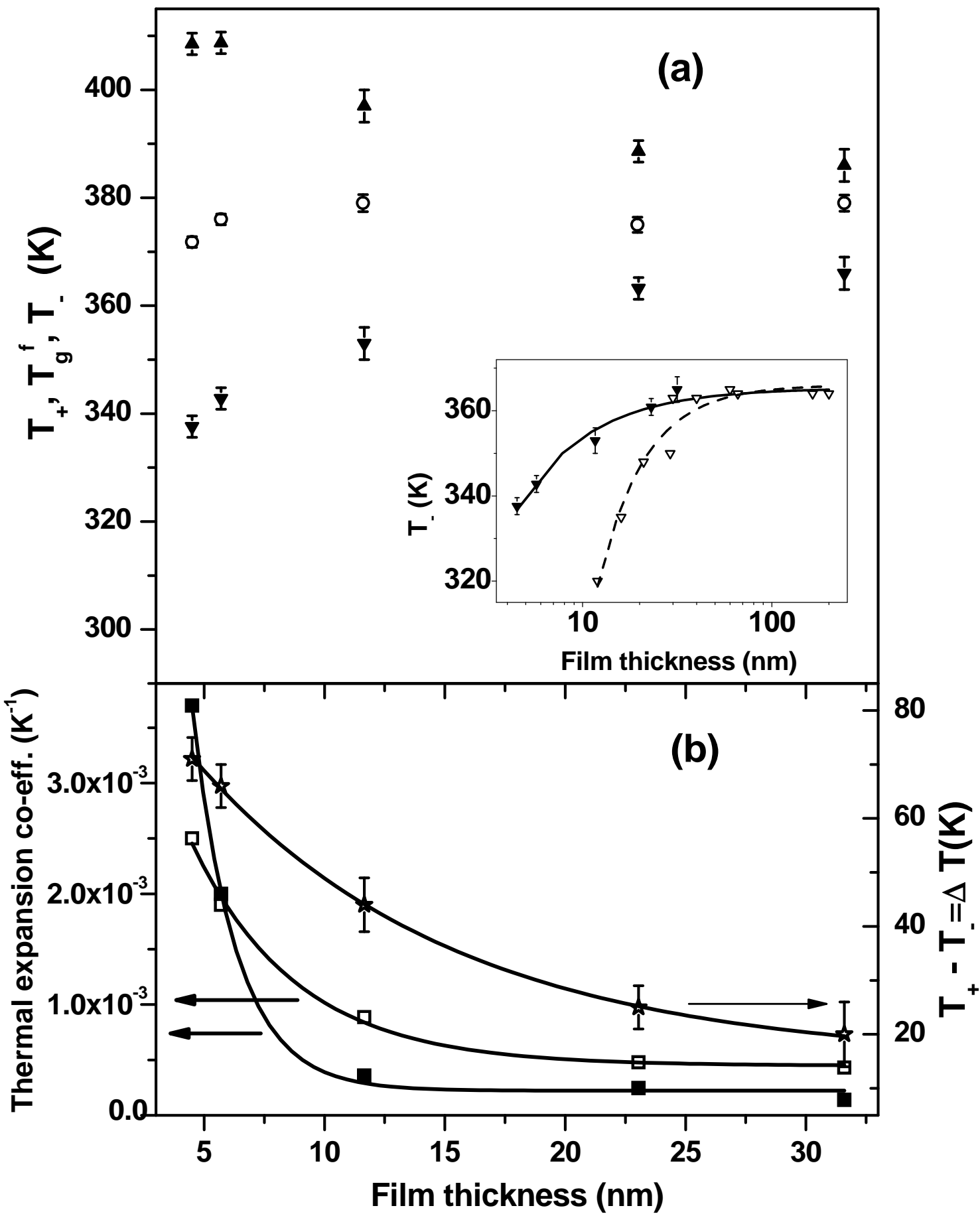
\end{document}